\documentclass[a4paper,fleqn,usenatbib]{mnras}

\usepackage{newtxtext,newtxmath}

\usepackage[T1]{fontenc}
\usepackage{ae,aecompl}

\usepackage{hyperref}
\usepackage{graphicx}	
\usepackage{url}
\usepackage{amsmath}	
\usepackage{amssymb}	
\usepackage{subfigure}
\usepackage{makecell}

\title[PSF DAE]{Point Spread Function Modelling for Wide Field  Small Aperture Telescopes with a Denoising Autoencoder}

\author[Peng Jia et al.]{
Peng Jia,$^{1,2,4}$\thanks{robinmartin20@gmail.com}
Xiyu Li,$^{1}$, Zhengyang Li $^3$, Weinan Wang$^{5}$ and Dongmei Cai$^{1}$\\
$^{1}$College of Physics and Optoelectronics, Taiyuan University of Technology, Taiyuan, 030024, China\\
$^{2}$Key Laboratory of Advanced Transducers and Intelligent Control Systems, Ministry of Education and Shanxi Province, \\Taiyuan University of Technology, Taiyuan, 030024, China\\
$^{3}$Nanjing Institute of Astronomical Optics and Technology CAS, Nanjing, Jiangsu, 210042, China\\
$^{4}$Department of Physics, Durham University, South Road, Durham, DH1 3LE, UK \\
$^{5}$Wuxi Internet of Innovation Center Company Limited, Wuxi, Jiangsu, 214135, China\\
}

\date{Accepted XXX. Received YYY; in original form ZZZ}

\pubyear{2015}

\begin{document}
\label{firstpage}
\pagerange{\pageref{firstpage}--\pageref{lastpage}}
\maketitle

\begin{abstract}
The point spread function reflects the state of an optical telescope and it is important for data post-processing methods design. For wide field small aperture telescopes, the point spread function is hard to model, because it is affected by many different effects and has strong temporal and spatial variations. In this paper, we propose to use the denoising autoencoder, a type of deep neural network, to model the point spread function of wide field small aperture telescopes. The denoising autoencoder  is a pure data based point spread function modelling method, which uses calibration data from real observations or numerical simulated results as point spread function templates. According to real observation conditions, different levels of random noise or aberrations are added to point spread function templates, making them as realizations of the point spread function, i.e., simulated star images. Then we train the denoising autoencoder with realizations and templates of the point spread function. After training, the denoising autoencoder learns the manifold space of the point spread function and can map any star images obtained by wide field small aperture telescopes directly to its point spread function, which could be used to design data post-processing or optical system alignment methods.
\end{abstract}




\section{Introduction}

Wide field small aperture telescopes (WFSATs) normally have a small aperture (around or less than 1 metre) and a wide field of view (several degrees). These properties make WFSATs light-weighted and low-cost. With remote control, WFSATs are widely used in optical observations for time domain astronomy \citep{burd2005pi,Ma2007,ping2017the,ratzloff2019building,sun2019precise}. Meanwhile since WFSATs are normally working automatically and no wavefront sensors are installed in them, they are hard to be maintained timely. Lack of maintenance would make quality of observation data change severely and limit their scientific output.\\

Aligning optical system remotely or increasing observation data quality with post-processing methods are two effective ways to increase the scientific output of WFSATs. For both of these two methods, the state of the whole optical system is required as prior knowledge. The point spread function (PSF) refers to the pulse response of the whole optical system and it can be used to describe states of a telescope \citep{racine1996telescope}. Several different PSF models have been proposed, such as analytical PSF modelling methods \citep{moffat1969theoretical} or data based PSF modelling methods \citep{jee2007principal}. \\

Analytical PSF modelling methods assume the PSF can be described by an analytical function with several experimental or physical parameters. The Moffat model is a widely used analytical PSF model which contains two parameters to describe the PSF. The Moffat model and basis functions based on Moffat model are candidate PSF reconstruction methods for general purpose survey telescopes \citep{Li2016}. For WFSATs, the Moffat model can fit the peak of star images, but it can not give promising results for the rest part. Because the field of view of WFSATs is very big, off-axis aberrations will bring highly deformable PSFs, which are hard to be described by circular symmetric functions \citep{piotrowski2013psf}.\\

Through careful analysis and  complicated computation, we can directly calculate PSFs of space-based telescopes with analytical PSF models and physical parameters \citep{rhodes2005modeling,rhodes2007stability,makidon2007jwst,perrin2014updated}. But directly computing PSFs of WFSATs is almost impossible, because WFSATs are seriously affected by complex off-axis aberrations, which are hard to be described or estimated by contemporary methods.\\

The principal components analysis (PCA) based PSF modelling method proposed by \citet{jee2007principal} is a pure data based  modelling method. It does not require complex calculations. If the number of star images is large enough and these images have adequate signal to noise ratio (SNR), the PCA based PSF modelling method can give promising results. Because WFSATs have larger field of views, shorter exposure time and smaller aperture size, many star images obtained by WFSATs have low SNR and low spatial sampling rate. In this circumstance, results obtained by the PCA based PSF modelling method are seriously affected by stars with different SNR \citep{wang2018automated} and we need a new PSF modelling method.\\

The autoencoder is a kind of deep neural networks, which can learn efficient data representation method under some regularization conditions. When linear activations are used, the optimal solution of an autoencoder is strongly related to the solution obtained by the PCA method \citep{bourlard1988auto}. With non-linear activations and different regularization conditions, autoencoders can obtain different data representations as required. The denoising autoencoder (DAE) is a special kind of autoencoder, which can obtain original data from distorted noisy data \citep{vincent2008extracting}. Because images obtained by WFSATs usually contain a lot of star images with low SNR, if we use the DAE method to replace the PCA method for PSF modelling,  we can use all star images as references in post-processing methods, which would increase robustness of these methods. In this paper, we will describe this DAE based PSF modelling method and discuss its possible applications. This paper is organized as follows: in section 2, we will introduce the DAE based PSF modelling method and compare it with the PCA based PSF modelling method. In section 3, we will test the DAE based PSF modelling method with simulated data and show how the DAE based PSF modelling method can increase the accuracy of secondary mirror alignment. In section 4, we make our conclusions and anticipate our future work.\\   

\section{Data based PSF modelling methods for WFSATs}

The quality of images obtained by optical telescopes is very sensitive to the outer environment. The atmospheric turbulence, temperature or gravity variation induced deformations will all introduce PSFs with temporal and spatial variations. According to our experience, PSFs of WFSATs are too complex to be modelled by analytical methods \citep{sun2017image}. Data based PSF modelling methods use statistical techniques to obtain PSFs from real observation data, which are elegant and do not need complex analysis of optical configurations of telescopes or fine-tuning of experimental parameters.\\ 

PCA is a widely used data based PSF modelling method. It is firstly proposed to model PSF of space-based telescopes \citep{jee2007principal} and later to model PSF of ground-based telescopes \citep{jee2011toward}. Right now, for general purpose sky survey telescopes with adequate spatial sampling rate and long enough exposure time, the PCA based PSF modelling method has been accepted as a standard method \citep{Bailey2012, Li2016}. \\

For WFSATs, which are generally used for fast all-sky survey, we have proposed to use the PCA based PSF modelling method to model the PSF and found that the PSF model can be used to increase astrometry accuracy \citep{sun2017image, Jia2017}. However, there are several drawbacks to use the PCA methods to model PSFs for WFSATs. First of all, WFSATs are a type of low cost telescopes and cameras installed in them have very small number of pixels (a star image with moderate SNR normally has around $3\times3$ to $5\times 5$ pixels). The low spatial sampling rate will reduce the number of effective components obtained by the PCA method. Secondly, because WFSATs have smaller aperture and shorter exposure time, very few stars have adequate SNR to be used as references. Our previous work shows that star images with different SNRs will lead to different results for the PCA based PSF modelling method \citep{wang2018automated}. Besides, if we only select star images with adequate SNR, the number of them would be too small and they will not distribute uniformly in the field of view. These problems will make the manifold space of PSFs obtained by PCA methods sub-optimal \citep{vidal2005generalized}.\\

It is commonly accepted that neural networks can be used to build an equivalent representation as that built by the PCA method \citep{bourlard1988auto}. Besides, the neural network has flexibility that we can add regularization condition to further increase its ability in representing data for different purposes. DAE is type of neural networks, which can map corrupted images to their uncorrupted version, according to the low-dimensional manifold of the training set. For DAE-based PSF modelling method, the low-dimension manifold is equivalent to the principal component space in the PCA method, albeit it is obtained by a slightly different way. The manifold of PSFs in WFSATs is built through training of DAE with pairs of real observation images and calibration images. After training, the DAE can map star images to their PSFs directly. We will discuss these two data-based PSF models in this section: the PCA model in subsection \ref{sec:PCA} and the DAE model in subsection \ref{sec:DAE}.\\

\subsection{PCA based PSF modelling method}
\label{sec:PCA} 

The PCA method was proposed in 1933 \citep{hotelling1933analysis}. It is a multivariate statistical technique which reduces the dimension of original data set to its low dimensional representation called principal components. In \cite{wang2018automated}, we further develop the traditional PCA based PSF model method \citep{jee2007principal} and propose a PCA based PSF model for WFSATs. Our method firstly uses the PCA method to obtain principle components as PSF basis and then uses self-organizing map (SOM) \citep{kohonen1982self} to cluster these PSFs according to their basis. We will briefly describe our method below.\\

We  obtain several star images $x_i$ from observation data as realizations of PSF and stretch all these images to vectors. These vectors will be placed in a matrix $x$ as shown in equation \ref{eq:eq1}. The size of $x$ is $N\times M$, where $N$ is the size of star images and $M$ is the number of star images. 
\begin{equation}
 \label{eq:eq1}
 x = [x_1,x_2, \cdots x_i]^T, \quad i= 1,\cdots,N
\end{equation}
Then, we will standardize vectors $x_i$ with equation \ref{eq:eq2} and \ref{eq:equation3}.
\begin{equation} 
  \label{eq:eq2}
mean= \frac{1}{N}\sum_{i=1}^{N}x_i
\end{equation}
\begin{equation} 
  \label{eq:equation3}
w_i =x_i-mean
\end{equation}
We use the singular value decomposition algorithm (SVD) to calculate eigenvalues $\lambda_i$ and eigenvectors $e_i$ of the covariance matrix $\Sigma$ as shown in equation \ref{eq:eq4}, where $W$ is a matrix composed of the column vectors $w_i$ placed side by side.
\begin{equation}  
  \label{eq:eq4}
\Sigma = WW^T 
\end{equation}
We sort eigenvalues according to their values and select the largest $K$ eigenvalues as effective components. The corresponding $K$ eigenvectors are basis of PSFs. With these eigenvectors, we can transform all star images into a new space $\Omega$ which has $K$ feature vectors as shown in equation \ref{eq:eq5}, where $y_i={e^T}_i w_i$.
\begin{equation}  
  \label{eq:eq5}
\Omega = \left[y_1 y_2 \cdots y_{{K}} \right]^T
\end{equation}
After PCA decomposition, star images are transformed to the PSF manifold space which has much fewer dimensions. We can then classify these PSFs in this space with SOM.\\

The SOM is an unsupervised competitive learning neural network, which mainly consists of an input layer and a competition layer. A node weight vector $m_i\in R^n$ connects with every node $i$ in the map, as shown in equation \ref{eq:eq6}.
\begin{equation} 
  \label{eq:eq6}
R^n = [m_1,m_2,\cdots,m_i]^T
\end{equation}
Weight vectors $m_i$ in different nodes will be firstly initialized by random number and then we will calculate the distance between each PSFs and node weights. The neuron with the smallest distance wins the competition and is set as the winning neuron $c$, as shown in equation \ref{eq:eq7}.
\begin{equation} \label{eq:eq7}
\left \|y- m_c\right \| = \min\limits_{i} {\left \|y- m_i  \right \|}
\end{equation}
$y$ is mapped onto the wining neuron $c$. After selecting the winner node, we will update weights of winning neuron's neighbours as defined in equation \ref{eq:eqe}.
\begin{equation} 
  \label{eq:eqe}
m_i(t+1) = m_i(t)+ h_{c,i}(t)[y(t)-m_i(t)]
\end{equation}
$t$ is the current iteration number and $h_{c,i}$ is the function to define weights of  neighbourhood neurons. The SOM repeats the process above in several iterations until $t$ becomes the maximal iteration number $t_{max}$ (in general, we set $t_{max}$ is 200). Finally, the network will classify PSFs into different clusters according to their relations to different nodes. We will then calculate mean PSF of all PSFs in the same cluster and use it as PSF of that area. Since the PCA based PSF modelling method is a statistical method, the effectiveness of this method strongly depends on the data amount and variety. Star images obtained by WFSATs normally have low SNR and it would introduce strong bias to the final results, if we only select stars with adequate SNR as references. \\

\subsection{DAE based PSF modelling method}

\label{sec:DAE} 
The autoencoder is a special kind of neural network, which has an encoder and a decoder. It compresses (encodes) the input data into data set with reduced dimension and reconstructs (decodes) the compressed data back to its original form. Through the encoding and the decoding process, the DAE is effective to learn the manifold space from the original data. \\

However there are some risks that the autoencoder will eventually become a "identity function", which simply learns a null function. A null function will output the input data directly and is not useful for our applications. In order to avoid this problem, it is necessary to add certain constraints. The denoising autoencoder \citep{vincent2008extracting,vincent2010stacked} is proposed to learn map between corrupted images and original images. The DAE has the same structure as that of the autoencoder, except that it adds different levels of noise to the input data during training. After training, the DAE learns a robust expression of manifold space of the input data \citep{vincent2010stacked,cha2019transformation}.\\ 

In this paper, we assume PSFs of WFSATs distribute in a manifold space that can be represented by calibration data from real observations, simulated data from physical calculations or mixture of them. PSFs represented by these data are called PSF templates. According to real observation conditions, we add different levels of noise or random aberrations to PSF templates to generate realizations of PSFs (simulated real observation star images). Then we train the DAE with PSF templates and realizations of PSFs. After training, the DAE is able to map real observation images to their original PSFs. Steps of our DAE based PSF modelling method are described below.\\ 

We extract star images $x_i$ with size of  $d \times d$  as input of the DAE. Their brightest pixel is in the centre of these images. Considering in real applications, there may exist error brought by the centroid algorithm, we set 1 pixel uncertainty in the training set and test set to increase the generalization ability of our neural network. Then we normalize star images with flux normalization algorithm as shown in equation \ref{eq:qua8},
\begin{equation} 
   p_i=\frac{x_i}{sum(x_i)}.
   \label{eq:qua8}
\end{equation}

In real applications, star images with different SNR may be used to restore their PSFs. To increase the generalization ability of the DAE, we use star images with different levels of SNR as the training set. We also find that the DAE is robust to the SNR and therefore we do not need to subtract the background before the flux normalization step.\\

\begin{figure}
\centering
\includegraphics[width=6cm,height=15cm]{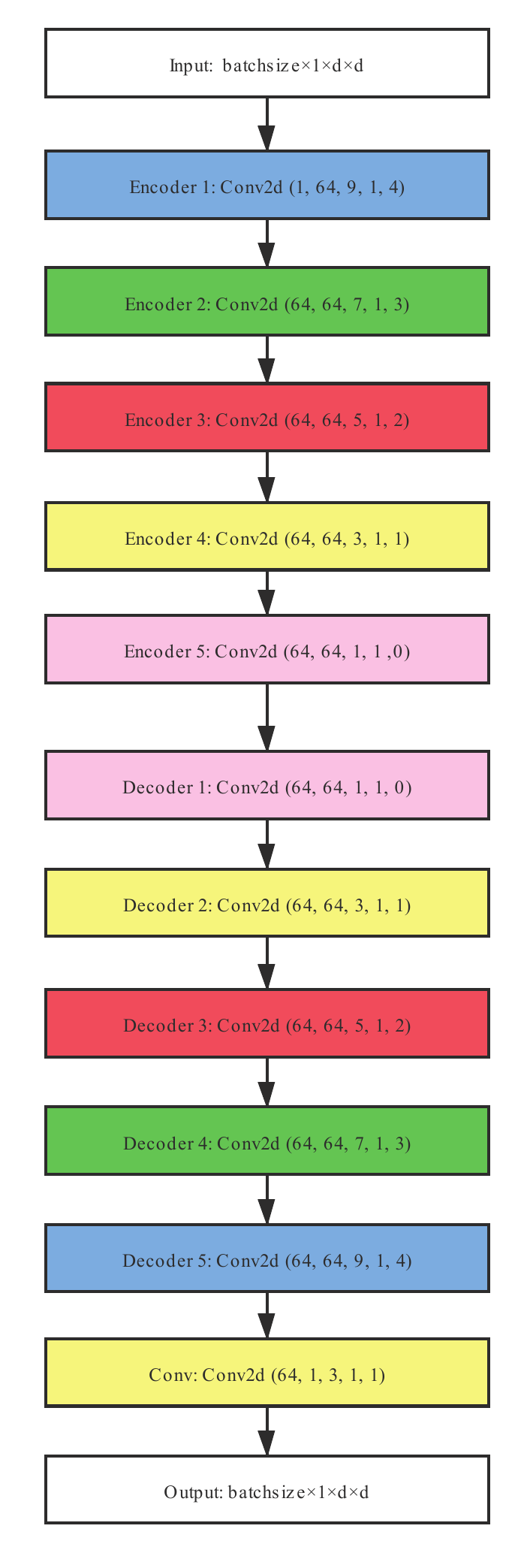}
\caption{Architecture of the DAE based PSF model. The box with Encoder stands for encoder layers. The box with Decoder stands for decoder layers. The Conv2d stands for the structure of that layer is a 2d convolution layer. Boxes with the same colour means convolutional kernel in that layer has the same size. The white boxes stand for the input and output layers.}
\label{fig:dae_struction}
\end{figure}

Normalized star images $p_i$ are input into the DAE as shown in figure \ref{fig:dae_struction}. We use conovolutional layers to build DAE in this paper, because convolutional layer is effective in building model with spatial connectivity \citep{cavallari2018unsupervised}. Our DAE contains 5 conovolutional layers as encoder and 5 convolutional layers as decoder \citep{ichimura2018spatial}. Each convolutional layer employs Rectified Linear Units (ReLU) as non-linear activation function. The convolutional kernels of the encoder or the decoder are organized in an inverted pyramid way. For the encoder, the kernel size is set as $9\times9$, $7\times7$, $5\times5$, $3\times3$ and $1\times 1$ respectively and for the decoder the kernel size is set as $1\times 1$, $3\times3$, $5\times5$, $7\times7$ and $9\times9$ respectively. With this structure, the convolutional layer uses a larger perceptual domain when it is closer to the input or output layer and vice versa.\\

We do not use pooling or unpooling layers in the DAE because the pooling function may discard useful details that are essential for PSF modelling. We only pad the input image to make the input image and the output image the same size. An input images $p_i$ is transferred through the DAE with the following steps. First of all, the autoencoder maps $p_i$ to its hidden representation $z_i$ which has much smaller dimensions $d' \times d'$, as shown in equation \ref{eq:quadratic},
\begin{equation}
   z_i = s(W\cdot{p_i}+b),
  \label{eq:quadratic}
\end{equation}
where $s$ is ReLU function. $z_i$ is then mapped back (decode) to $y_i$, which has the same size as $x_i$. With size of $d\times d$, $y_i$ can be viewed as the reconstructed PSF.
\begin{equation}
  y_i=s(W^{'} \cdot z_i+b^{'})
  \label{eq:eq8}
\end{equation}
$W$ and $W'$ are weight matrix with size of $d'\times d$ and $d \times d'$ respectively. $b$ and $b'$ are bias matrix with size of $d'\times d'$ and $d\times d$ respectively. These parameters $(W,W',b,b')$ are optimized to minimize reconstruction error, which can be assessed by different loss functions such as mean squared error or cross-entropy.

\begin{equation}
\label{eq:mse}
L_M(p_i,y_i)= \sum_{i=1}^N {\lVert{p_i-y_i}\rVert}^2
\end{equation}

\begin{equation}
\begin{aligned}
\label{eq:cross_entropy}
L_H(p_i,y_i)  = & H(B_{p_i}||B_{y_i}) \\
                  = & -\sum_{k=1}^d[{{p_i}_k}\log_{{y_i}_k}+(1-{p_i}_k)\log_{(1-{y_i}_k)}]
\end{aligned}
\end{equation}

$L_M$ is the traditional mean squared error. $L_H$ stands for the cross-entropy, which assumed $p_i$ and $y_i$ as matrix of bit probabilities, and ${p_i}_k$ or ${y_i}_k$ is normalized star images and its corresponding PSF.  In this paper, we use the $L_M(p_i,y_i)$ as loss function.\\

\section{Applications of the DAE based PSF Model}

In this part, we test the performance of DAE based PSF modelling method with simulated data. There are three scenes in this part. The first scene is modelling PSF for a telescope with field dependent aberrations and the second scene is modelling PSF for a telescope with atmospheric turbulence induced static aberrations. The first and the second scene are used to show that the DAE based PSF model is capable to learn effective PSF representation, even for images with low SNR or highly variable PSFs. In the third scene, we will show that our DAE based PSF modelling method can increase accuracy of the secondary mirror alignment algorithm.\\

The DAE based PSF Model is implemented by pytorch \citep{kossaifi2019tensorly} and CUDA \citep{grimm2015helios} in a computer with Intel Core E5-2620 v3 and NVIDIA Tesla K40 GPU. Hyper-parameters, such as the learning rate, epoch size and optimization method are important regularization conditions. In this paper, we set epoch = 100, batchsize = 125, learning rate = 0.00005. The Adam optimization algorithm \citep{kingma2014adam} is used for optimization with the MSE as loss function. We will discuss details of these three scenes below.\\

\subsection{Test the DAE based PSF Modelling method with a simulated wide field telescope}

\begin{figure}
	\centering
	\includegraphics[width=6cm]{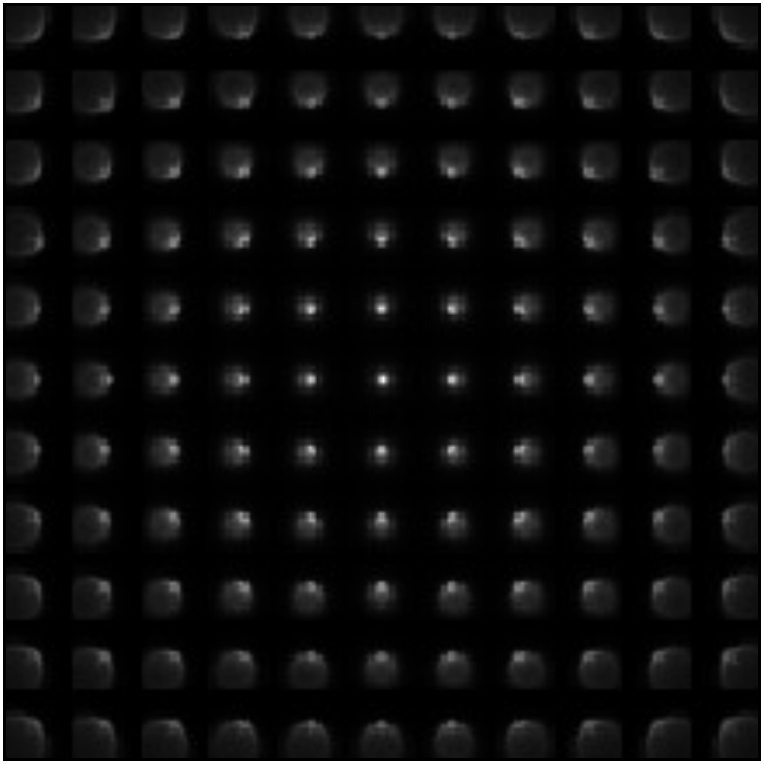}
	\caption{This figure shows shape and position of 121 PSFs in the full field of view.}
	\label{fig:allview_data}
\end{figure}

 In this part, we simulate a WFSAT with parameters listed in table \ref{table:WFSAT}. It is a classical reflective telescope with small aberrations. However we add large field-dependent Seidel aberrations (coma and astigmatism) to its primary mirror to increase the spatial variability of its PSFs. We calculate 121 images with size of  $16\times 16$ pixels in the whole field of view through Fresnel propagation \citep{Perrin2016} and these images are separated by $1.1^{\circ}$ as shown in figure \ref{fig:allview_data}. Since no additional noise is added to these images, they can be viewed as PSF templates of this telescope. In this scene, we assume aberrations of this telescope are known and test whether the DAE based PSF model can obtain PSF from real noisy observation data. While in real applications, aberrations are usually unknown to users and PSF templates obtained from real observations would be better.\\

 \begin{table}
	\caption{Parameters of a simulated WFSAT for the first scene}
	\begin{tabular}{ccc}
		\hline
		Parameters  & Values \\
		\hline
		Optical Design &  Cassegrain telescope\\
		Aperture Diameter & 1.0 meter\\
		field of view & $12.1^{\circ} \times 12.1^{\circ} $\\
		pixelscale & 0.01 arcsec\\
		Spherical aberration & 0.500 wavelengths\\
		Coma & 4.000 wavelengths\\
		Field curvature & 1.813 wavelengths\\
		Astigmatism & 4.196 wavelengths\\
		Distortion & -0.113 wavelengths\\
		\hline 						
	\end{tabular}	
	\label{table:WFSAT}
\end{table}

Data regularization condition is important for the DAE. For our application, we use two methods to generate regularized data: adding different levels of random aberrations to its primary mirror and different levels of noise to change SNR of star images as shown in table \ref{table:aberrations and noise}. The Poisson distribution is used to simulate the photon noise and the background noise. Different levels of noise are added to each data set. $\lambda$ which is shown in table \ref{table:aberrations and noise} stands for the worst case and it will change inside the same data set to make star images have different levels of SNR. Different levels of random wavefront aberrations, represented by low order Zernike polynomials, are added to the primary mirror of this telescope to generate random interference, which would increase generalization ability of the DAE. The coefficients of these random aberrations are set as percentage of that of static Seidel coefficients. This simulation is close to real situations. During real observations, the atmospheric turbulence will introduce random aberrations and the noise of different level will affect SNR of observed images, while we need to obtain static aberrations represented by PSF templates from these observation data.\\

\begin{figure}
	\centering
	\includegraphics[width=\columnwidth]{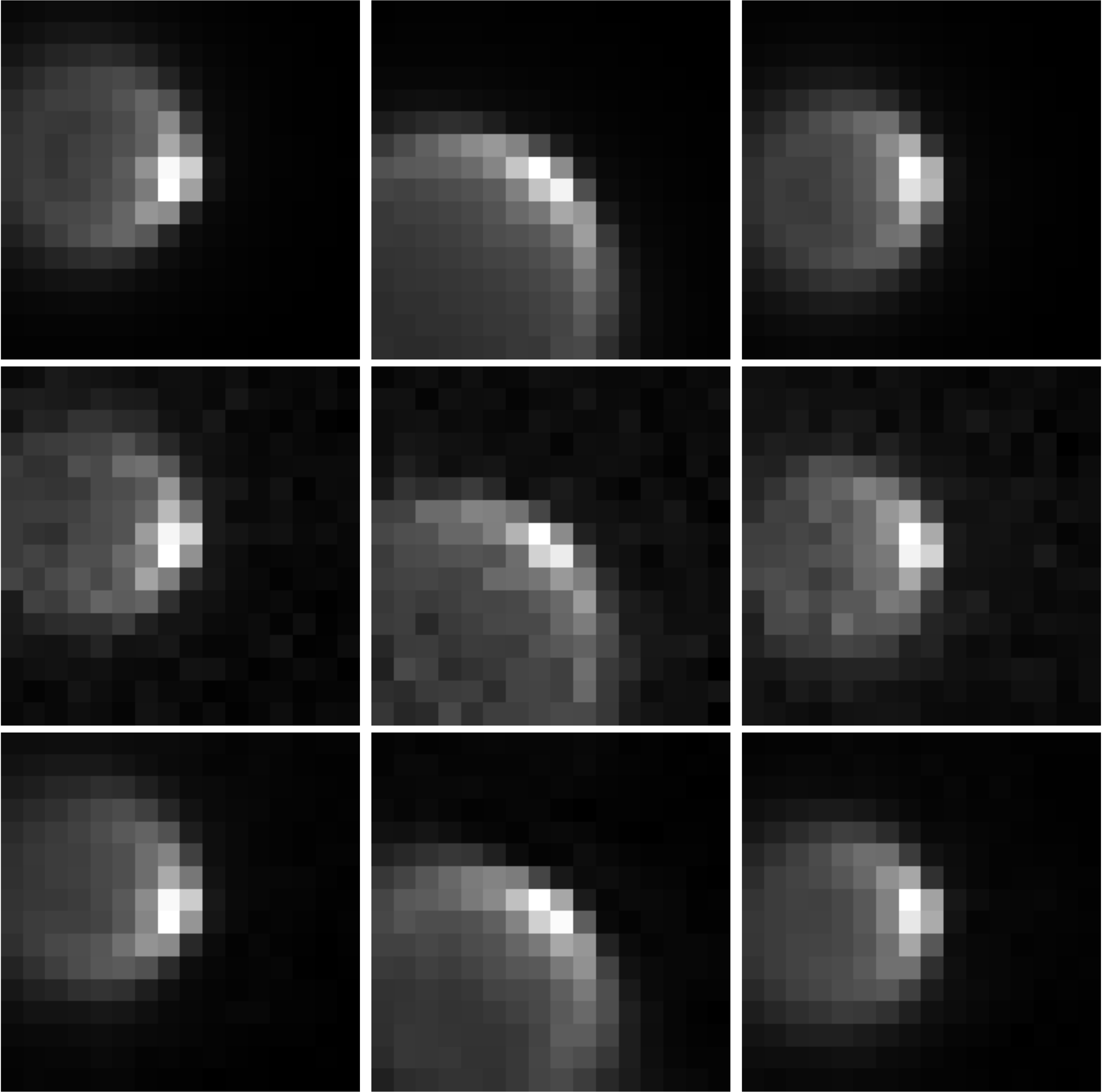}
	\caption{This figure shows results obtained by the DAE based PSF model for the dataset1. Images in the first row show original images, images in the second row show noisy images with $\sigma=0.0005$ and images in the third row show PSFs obtained by the DAE.}
	\label{fig:sigma0.0005}
\end{figure}

We firstly use star images with relatively high SNR from dataset1 to train the DAE based PSF model. We randomly pick 6724 star images as training set and 1681 star images as test set. After training, we use the DAE to obtain PSFs from star images in the test set. Several results are shown in figure \ref{fig:sigma0.0005}. From these figures, we can find that when the noise level is low, the DAE based PSF model is able to obtain original PSFs from star images directly.\\

\begin{figure}
	\centering
	\includegraphics[width=\columnwidth]{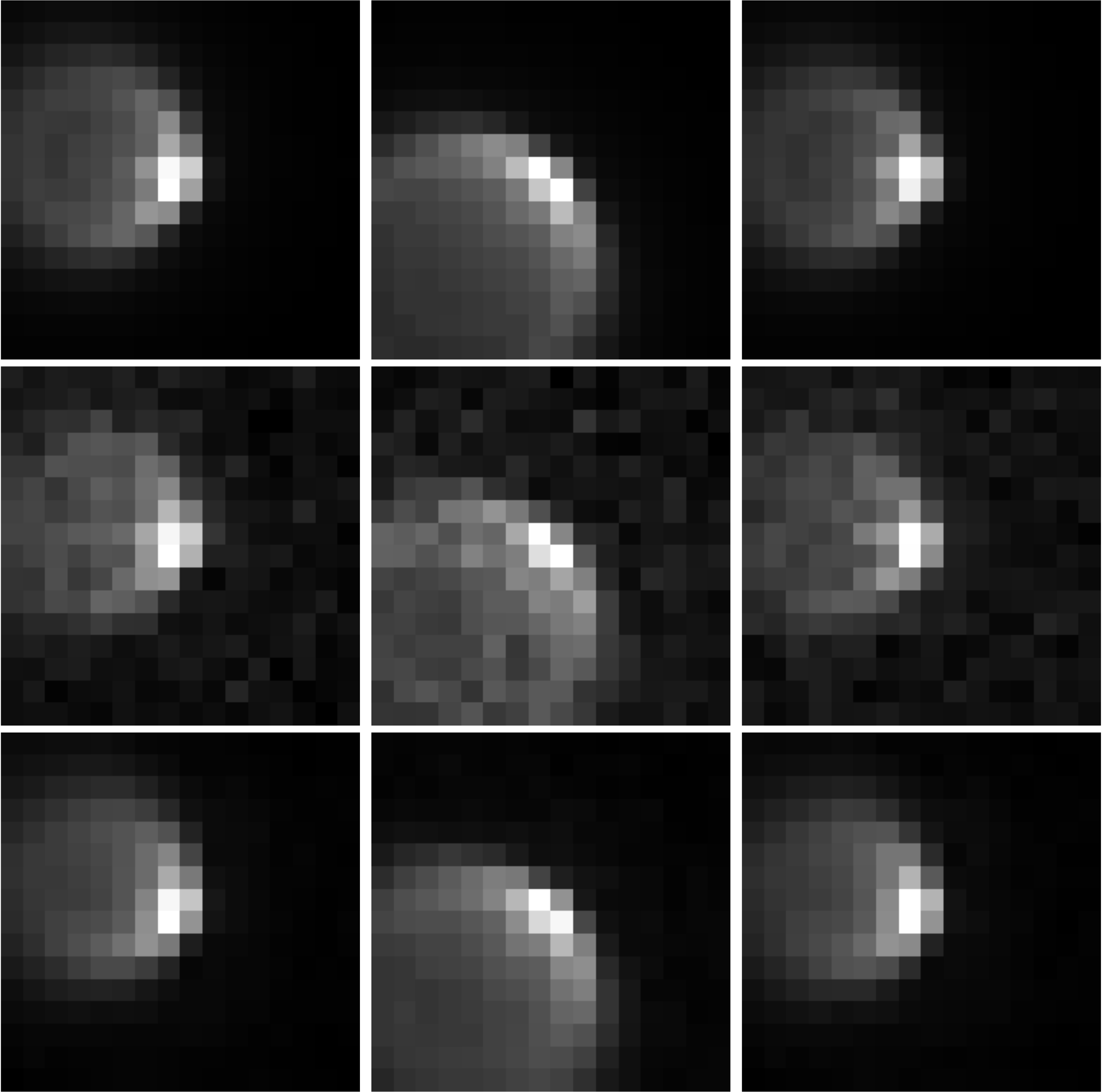}
	\caption{This figure shows results obtained by the DAE based PSF model for the dataset2. Images in the first row show original images, images in the second row show noisy images with $\sigma=0.001$ and images in the third row show PSFs obtained by the DAE.}
	\label{fig:sigma0.001}
\end{figure}

\begin{figure}
	\centering
	\includegraphics[width=\columnwidth]{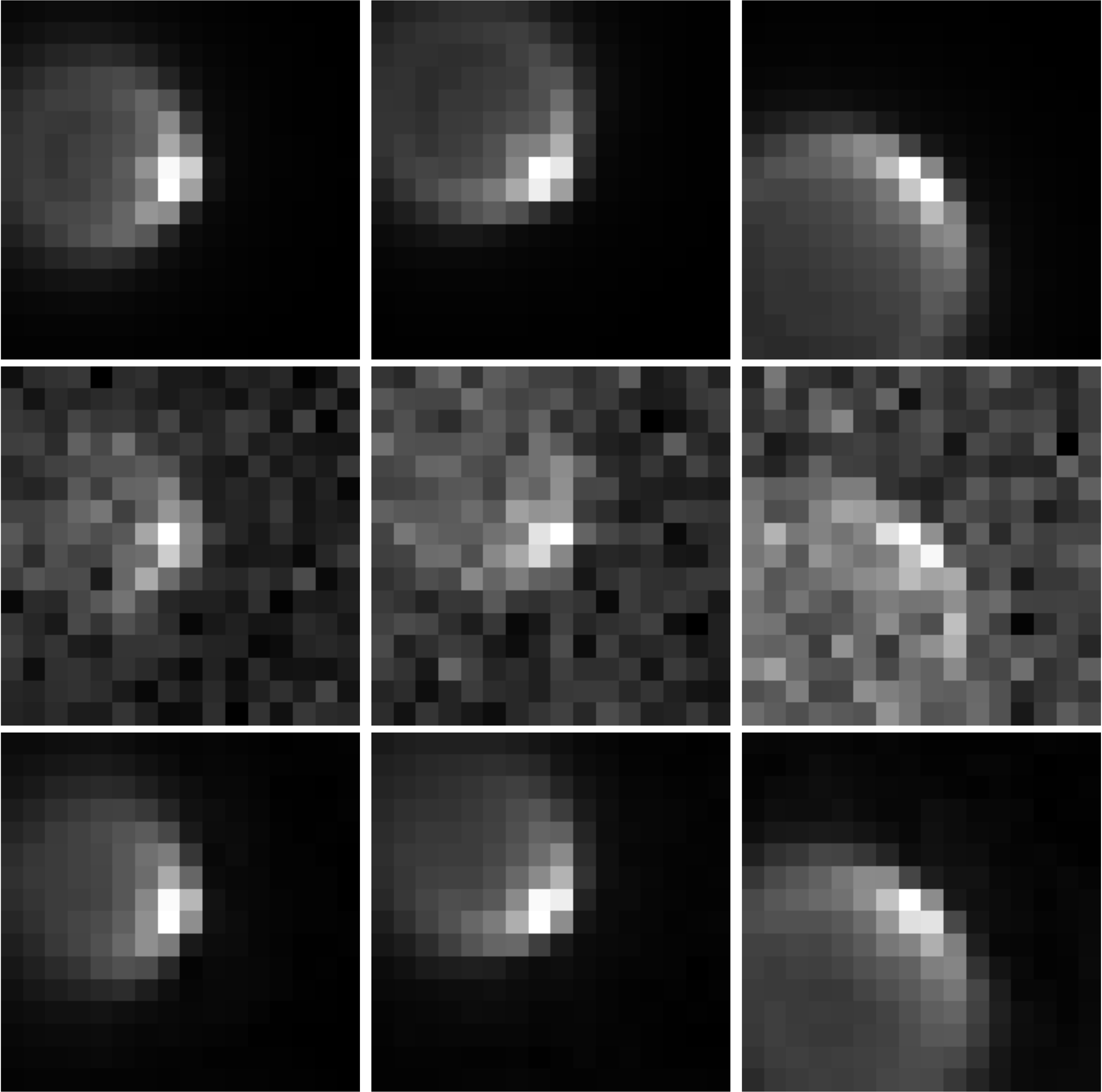}
	\caption{This figure shows results obtained by the DAE based PSF model for the dataset3. Images in the first row show original images, images in the second row show noisy images with $\sigma=0.003$ and images in the third row show PSFs obtained by the DAE.}
	\label{fig:sigma0.003}
\end{figure}

Then we use star images with slightly smaller SNR to test the DAE. We also pick 6724 star images randomly as training set and 1681 star images as test set from dataset2. The results are shown in figure \ref{fig:sigma0.001}. We can find that with larger noise level, the PSF obtained by the DAE is almost the same from the original PSF. We contiguously increase the noise level and generate images with lower SNR to test the DAE based PSF modelling method. We find that the DAE based PSF modelling method is robust. As shown in figure \ref{fig:sigma0.003}, when the noise level ($\lambda$) is 0.003, it is almost impossible for human beings to recognize the original PSF from star images. The DAE is still able to obtain the original PSF. We further use the SSIM (structural similarity index) and the MSE (Mean Squared Error) functions from the scikit-image package \citep{van2014} to evaluate PSFs reconstructed by the PCA based and the DAE based PSF modelling methods. As shown in table \ref{table:dataset3_ssim} and \ref{table:dataset3_MSE}, the DAE based PSF modelling method is able to achieve much higher SSIM and much smaller MSE.\\

\begin{figure}
	\centering
	\includegraphics[width=\columnwidth]{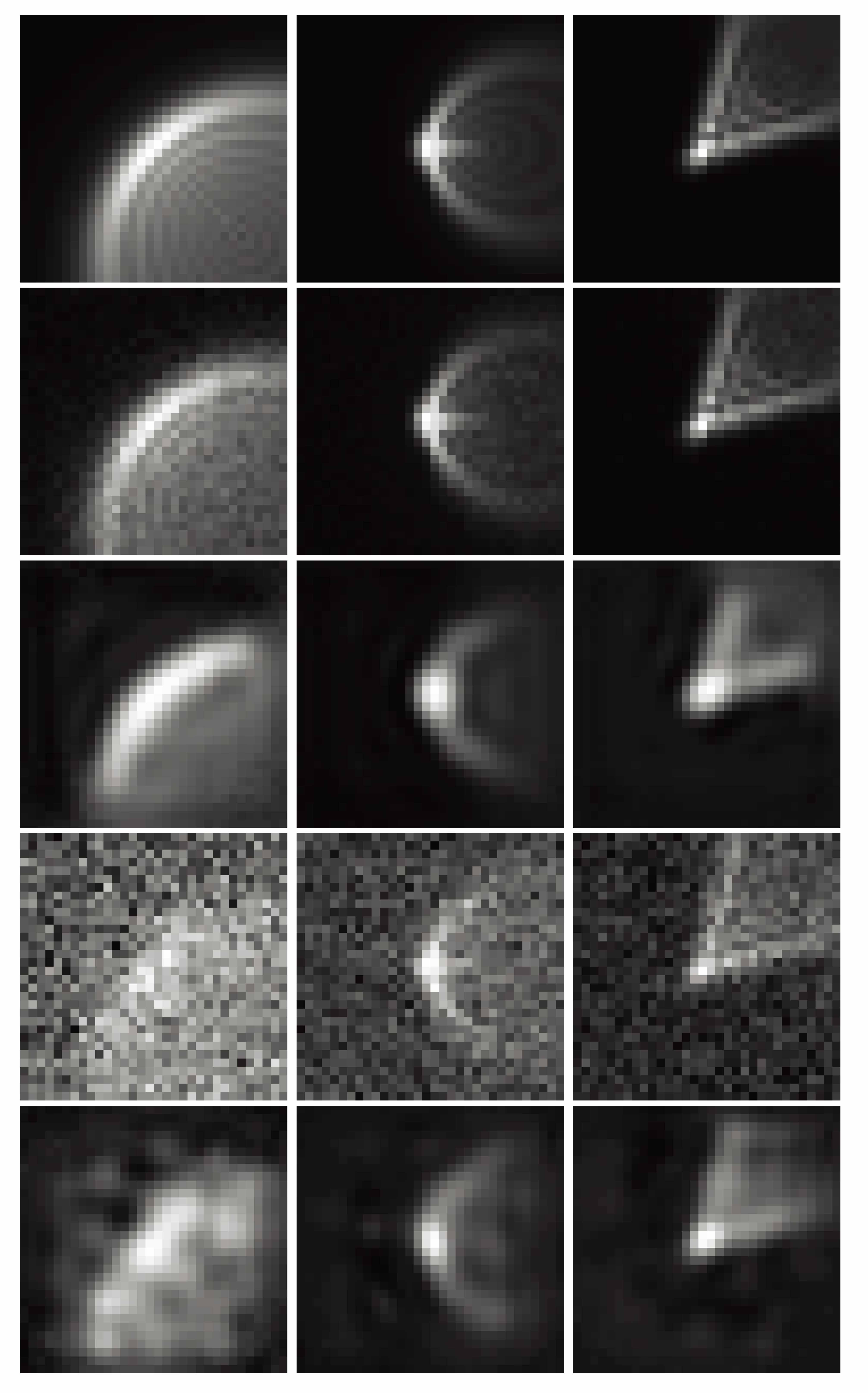}
	\caption{This figure shows results obtained by the DAE based PSF model for the dataset4 and the dataset5. Images in the first row show original images, images in the second row show noisy images with $\sigma=0.00003$ and images in the third row show PSFs obtained by the DAE. Images in the fourth row show noisy images with $\sigma=0.0009$, and images in the fifth row show PSFs obtained by the DAE from noisy images in the fourth row.}
	\label{fig:random0.5_noise}
\end{figure}

We also test the DAE based PSF modelling method with 50\% random aberrations and the results are shown in figure \ref{fig:random0.5_noise}. When the SNR is large, we can find that the original PSF can be obtained.  As shown in table  \ref{table:dataset5_ssim} and \ref{table:dataset5_MSE}, we also use the SSIM and the MSE to evaluate results obtained by the DAE based and the PCA based PSF modelling methods. We can find that the DAE based PSF modelling method can still achieve better performance with star images of high SNR, even when the random aberration is big.\\

However, when we reduce the SNR, the results obtained by the DAE based PSF modelling method are not consistent. PSFs obtained by star images at the centre of the field of view are relatively good, but PSFs obtained by star images at the edge of the field of view are not good. It is probably caused by the way we add random aberrations. Since we add wavefront aberration in percentage, at the edge of the field of view, when the aberration is larger, the random interference will be larger. Large random interference will make the DAE based PSF modelling method ineffective. Meanwhile, it also indicates us that the performance of DAE based PSF modelling method is limited by outer interference and random noise. When random aberration is larger than 50\% of its original aberrations and observed images are affected by large random noise, the DAE based PSF modelling method can not give promising results.\\

\begin{table}
	\caption{Simulated star images with different levels of noise and aberration. In Noise1, $\lambda$ = 0.00003, 0.00005, 0.00007 and 0.00009 respectively. In Noise2, $\lambda$ = 0.0003, 0.0005, 0.0007 and 0.0009 respectively, and in Noise3, $\lambda$ = 0.001, 0.002, 0.003 and 0.005 respectively. $\lambda$ is expectation of Poisson distribution. Random aberration with different percentage stands for random Seidel aberrations which satisfies normal distribution with zero mean and variance of different percentage of its original aberrations. Random aberration with different percentage stands for random Zernike low order aberrations which satisfies normal distribution with zero mean and variance of different percentage of its original aberrations.}
	\begin{tabular}{ccccc}
		\hline
		Aberration & Noise1 & Noise2 & Noise3\\ 
		\hline
		10\% random aberration &  dataset1 &  dataset2 & dataset3\\
		50\% random aberration &  dataset4 &  dataset5 & dataset6 \\
		\hline 						
	\end{tabular}	
	\label{table:aberrations and noise}
\end{table}

\begin{table}
	\caption{SSIM of dataset3.}
	\begin{tabular}{ccccc}
		\hline
		SSIM & $\sigma=0.001$ &  $\sigma=0.002$ &  $\sigma=0.003$ &  $\sigma=0.005$\\ 
		\hline
		original-noise &0.9512  &0.8347  &0.7326 &0.5258 \\
		PCA            &0.9237  &0.9242  &0.9220 &0.9147  \\
		DAE            &0.9549  &0.9548  &0.9544 & 0.9539 \\
		\hline 						
	\end{tabular}	
	\label{table:dataset3_ssim}
\end{table}

\begin{table}
	\caption{MSE of dataset3.}
	\setlength{\tabcolsep}{0.6mm}{
	\begin{tabular}{ccccc}
		\hline
		MSE & $\sigma=0.001$ &  $\sigma=0.002$ &  $\sigma=0.003$ &  $\sigma=0.005$\\ 
		\hline
		original-noise &$2.6600\times10^{-5}$   &$1.0957\times10^{-4}$   &$2.1422\times10^{-4}$  &$6.0358\times10^{-4}$ \\
		PCA            &$3.5375\times10^{-5}$  &$3.4552\times10^{-5}$  &$3.8546\times10^{-5}$   &$5.2028\times10^{-5}$  \\
		DAE            &$2.7568\times10^{-5}$  &$2.7631\times10^{-5}$  &$2.7842\times10^{-5}$ & $2.8271\times10^{-5}$ \\
		\hline 						
	\end{tabular}	}
	\label{table:dataset3_MSE}
\end{table}

\begin{table}
	\caption{SSIM of dataset5.}	
	\begin{tabular}{ccccc}
		\hline
		SSIM & $\sigma=0.0003$ &  $\sigma=0.0005$ &  $\sigma=0.0007$ &  $\sigma=0.0009$\\ 
		\hline
		original-noise &0.9949  &0.9853  &0.9728 &0.9562 \\
		PCA            &0.9935  &0.9937  &0.9936 &0.9934 \\
		DAE            &0.9993  &0.9993  &0.9993 & 0.9992 \\
		\hline 						
	\end{tabular}	
	\label{table:dataset5_ssim}
\end{table}

\begin{table}
	\caption{MSE of dataset5.}
	\setlength{\tabcolsep}{0.6mm}{
	\begin{tabular}{ccccc}
		\hline
		MSE & $\sigma=0.0003$ &  $\sigma=0.0005$ &  $\sigma=0.0007$ &  $\sigma=0.0009$\\ 
		\hline
		original-noise &$2.1611\times10^{-6}$   &$6.3513\times10^{-6}$   &$1.1926\times10^{-5}$  &$1.9634\times10^{-5}$ \\
		PCA            &$2.5760\times10^{-6}$  &$2.5124\times10^{-6}$  &$2.6983\times10^{-6}$   &$2.9674\times10^{-6}$  \\
		DAE            &$3.7504\times10^{-7}$  &$3.8568\times10^{-7}$  &$3.9869\times10^{-7}$ & $4.4674\times10^{-7}$ \\
		\hline 						
	\end{tabular}	}
	\label{table:dataset5_MSE}
\end{table}

\subsection{Test the DAE based PSF Model with a simulated telescope affected by static atmospheric turbulence aberrations}
 In this part, we consider a telescope with more complex aberrations. It is an ideal telescope with static atmospheric turbulence induced wavefront aberrations in its pupil. In this scene, PSFs would have highly spatial variation. We use this scene to test the performance of the DAE in modelling complex PSFs.\\

\begin{figure}
	\flushleft
	\includegraphics[width= \columnwidth]{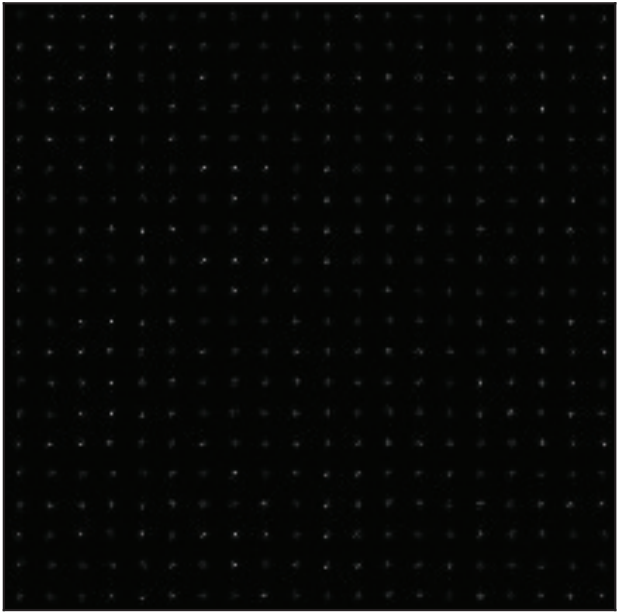}
	\caption{This figure shows PSFs for an aberrations free telescope with static atmospheric turbulence aberrations. There are 400 PSFs distributed equally in a field of view of 14 arcmin.}
	\label{fig:short_allview}
\end{figure}

We generated PSF templates with size of $24 \times 24$ pixels in 400 locations equally distributed in a field of view of 14 arcmin, as shown in the figure \ref{fig:short_allview}. The atmospheric turbulence phase screen is generated by the method proposed in \citet{jia2015simulation,jia2015real} and we use the Durham Adaptive Optics Simulation Platform to generate PSFs  \citep{Basden2018}. We add different levels of noise to PSFs to make them as simulated star images in dataset7 and dataset8. In dataset7, Poisson noise is added to PSFs with $\lambda = 0.0003$, 0.0005, 0.0007 and 0.0009. In  dataset8, Poisson noise with $\lambda$ = 0.001, 0.002, 0.003 and 0.005 is added to these PSFs. The $\lambda$ used in this part is the same as we defined in previous section: it stands for the worst case in each dataset.\\

\begin{figure}
	\centering
	\includegraphics[width=\columnwidth]{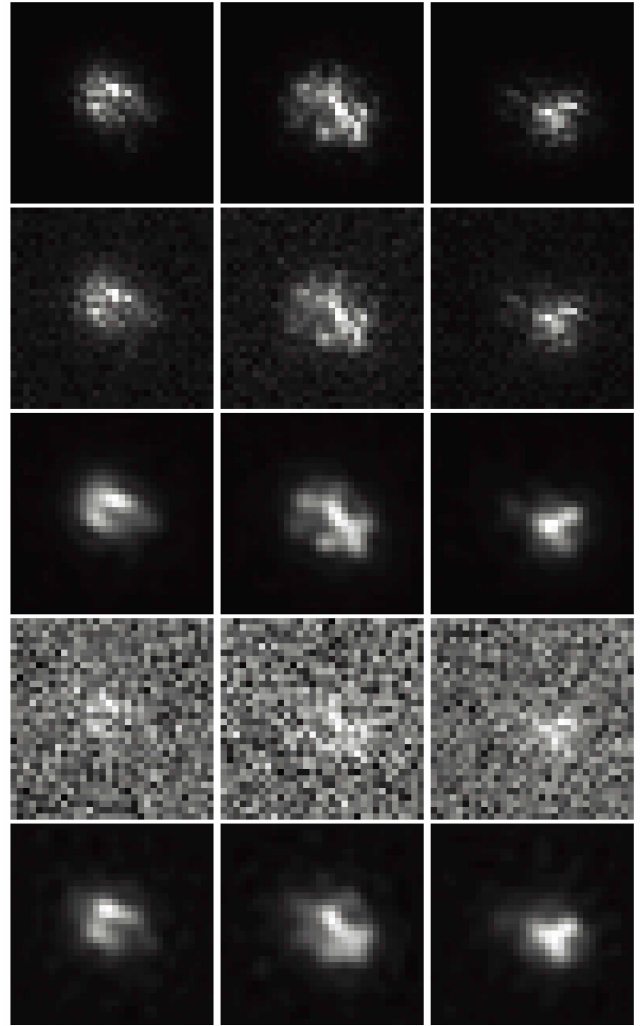}
	\caption{This figure shows the results obtained by the DAE PSF modelling method in dataset7 and dataset8. Images in the first row are original PSFs, images in the second row show noisy images with $\lambda=0.0005$, and images in the third row show PSFs obtained by the DAE based PSF model from noisy images in the second row. Images in the fourth row show noisy images with $\lambda=0.005$, and images in the fifth row are PSFs obtained by the DAE based PSF model from noisy images in the fourth row.}
	\label{fig:short_bignoise}
\end{figure}

We use star images from dataset7 or dataset8 to train two DAEs. We randomly pick 8000 star images as training set and 2000 star images as test set for each of these DAEs. After training, we use the trained DAE to obtain PSFs from star images in the test set. We find that the DAE is robust when $\lambda$=0.0005, as shown in figure \ref{fig:short_bignoise}. When $\lambda$=0.005, it is almost impossible for human beings to recognize original PSFs from star images, the DAE based PSF model can still obtain part of original PSFs. These tests show that DAE based PSF modelling method has a relatively good representation ability in modelling PSF with complex structure. Although when SNR is extremely low, its performance will drop down. We also use the SSIM and the MSE to evaluate performance of the DAE based and the PCA based PSF modelling methods. As shown in table \ref{table:dataset8_ssim} and \ref{table:dataset8_MSE}, we can find that the DAE based PSF modelling method has better performance than the PCA based PSF modelling method.\\

\begin{table}
	\caption{SSIM of dataset8.}
	\begin{tabular}{ccccc}
		\hline
		SSIM & $\sigma=0.001$ &  $\sigma=0.002$ &  $\sigma=0.003$ &  $\sigma=0.005$\\ 
		\hline
		original-noise &0.9430  &0.7612  &0.6866 &0.4288 \\
		PCA            &0.9874  &0.9858  &0.9832 &0.9749\\
		DAE            &0.9995  &0.9994  &0.9994 &0.9992  \\
		\hline 						
	\end{tabular}	
	\label{table:dataset8_ssim}
\end{table}

\begin{table}
	\caption{MSE of dataset8.}
	\setlength{\tabcolsep}{0.6mm}{
	\begin{tabular}{ccccc}
		\hline
		MSE & $\sigma=0.001$ &  $\sigma=0.002$ &  $\sigma=0.003$ &  $\sigma=0.005$\\ 
		\hline
		original-noise &$2.6071\times10^{-5}$   &$1.3834\times10^{-4}$   &$2.0626\times10^{-4}$  &$6.2944\times10^{-4}$ \\
		PCA            &$5.7382\times10^{-6}$  &$8.2983\times10^{-6}$  &$1.2569\times10^{-5}$   &$2.6235\times10^{-5}$  \\
		DAE            &$7.1009\times10^{-7}$  &$7.6685\times10^{-7}$  &$8.4012\times10^{-7}$ & $1.0136\times10^{-6}$ \\
		\hline 						
	\end{tabular}}	
	\label{table:dataset8_MSE}
\end{table}

\subsection{Secondary Mirror Alignment with a Convolutional Neural Network and DAE PSF model}
To better show increments brought by our DAE based PSF model to other post-processing or telescope alignment methods, we consider a real application case in this subsection. Secondary mirror alignment is a common problem for real observations in wide field survey telescopes, because these telescopes normally have small F-number and the performance of these telescopes is very sensitive to secondary mirror mis-alignment \citep{Li2015}.\\

For secondary mirror alignment, astronomers need to obtain the position of the secondary mirror. We consider four degrees of freedom for the secondary mirror in this paper: decenter along $X$ and $Y$ directions and tilt along $X$ and $Y$ directions. Because misalignment will introduce PSF variations in the whole field of view, we can obtain the amount of misalignment according to PSFs in different field of views. Obtaining the amount of misalignment according to variation of PSFs is a traditional regression problem and it can be solved through machine learning techniques. It should be noted that we set the CCD plane in a fixed position and do not consider decenter along $Z$ direction, because these two degrees of freedom are highly correlated and are hard to be directly solved by a machine learning algorithm.\\

\begin{figure}
	\centering
	\includegraphics[width=\columnwidth]{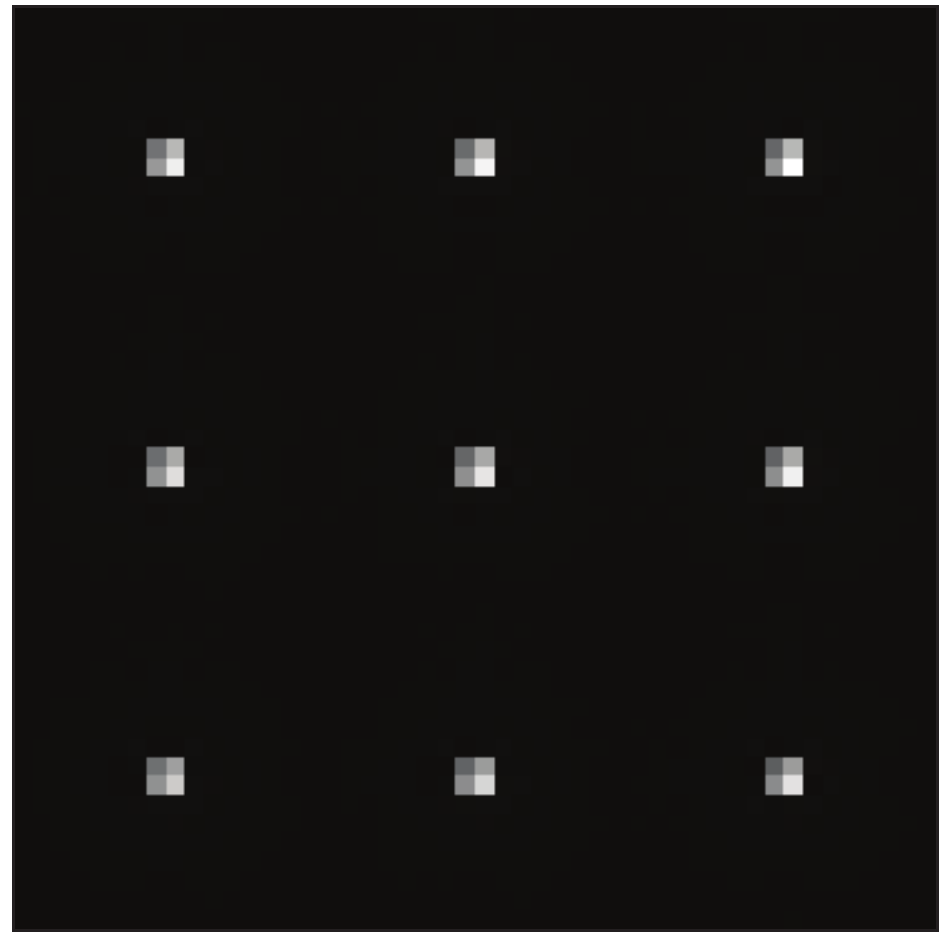}
	\caption{This figure shows distribution of star images that are used for secondary mirror alignment. They are distributed in the centre and 8 corners of the field of view.}
	\label{fig:9psf}
\end{figure}

\begin{figure}
	\centering
	\includegraphics[width=\columnwidth]{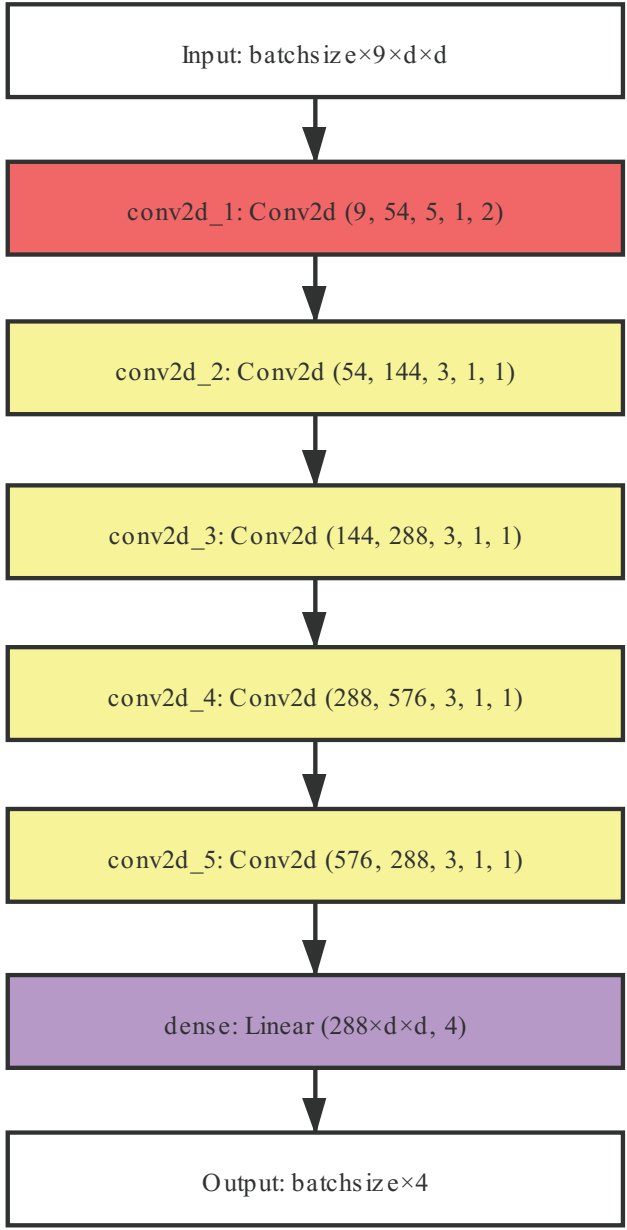}
	\caption{This figure shows the structure of the CNN used for regression of secondary mirror alignment. The input dimension is $9 \times d \times d$. $d$ and $d$ are the width and height of the image respectively. In the figure, Conv2d represents the convolutional layer and Linear represents the full connection layer.}
	\label{fig:logist}
\end{figure}

In this paper, we consider a Ritchey–Chrétien telescope with a field corrector, which is adapted from a sample file in Zemax. The telescope has a diameter of 1.5 metre and a field of view of 1 degree. We use 9 PSFs obtained from centre and corners of the field of view to obtain the mount of misalignment as shown in figure \ref{fig:9psf}. A simple convolutional neural network (CNN) is proposed in this paper to solve the regression problem and the structure of this CNN is shown in figure \ref{fig:logist}. There are five convolution layers and a fully connected layer in this CNN. We use batch normalization \citep{ioffe2015batch} after each convolution layer and select Leaky--ReLU function \citep{laurent2017multilinear} as activation function. We use original PSFs (images with 9 channels and in each channel is the PSF in different position) as input and the amount of misalignment (4 dimensions and they stand for decenter along the x and y directions and tilt along the x and y directions) as output to train the CNN. The CNN is trained with batchsize=10 and epoch=100. The learning rate is 0.001 at the begining and we update the learning rate after 30 epochs with equation \ref{eq:lr},
\begin{equation}
\label{eq:lr}
lr=0.001*(0.1^{\left \lfloor epoch\/30 \right \rfloor}),
\end{equation}
where $epoch$ is the epoch number, $\left \lfloor \right \rfloor$ stands for the floor function. We use Adams algorithm with MSE loss function to update weights in the CNN. After training, the CNN can output the value of decenter and tilt along $X$ and $Y$ directions directly according to 9 PSFs.\\

The amount of misalignment lies between -0.1 to 0.1 degree for tilt and -0.1 to 0.1 centimetre for decenter. We obtain Zernike coefficients for different field of views through continuously adjust the amount of misalignment. Then we calculate PSFs according to the Zernike coefficients through Fresnel propagation \citep{Perrin2016}. We obtain 625 states of misalignment and there are 9 PSFs in each state. We add Poisson noise with $\lambda$ of 0.002 and 0.005 to these PSFs to make simulated observation images. $\lambda$ used in this part is the same as we defined in above two subsection: they stand for the worst case of each dataset. We use 5625 simulated PSFs to train the DAE PSF model with steps discussed in the start of Section 3. After training, the DAE PSF model can output PSFs directly according to observation images.\\

We generate a new set of observation images with misalignments in the same range and noise within the same level as test set. We firstly input the test set into the CNN to obtain the amount of misalignment directly. The results obtained by this way stand for a common situation of secondary mirror alignment, where we directly use a trained CNN to obtain the amount of mis-alignment without considering the PSF model. Meanwhile, we input the test set into the DAE based PSF model to obtain PSFs and input these PSFs into the CNN to obtain the mount of misalignment. The results are shown in table \ref{tab:diff_0.002} and \ref{tab:diff_0.005}. As can be seen from these tables, the CNN is robust to noise level, if we use it for misalignment estimation. It can give relatively good estimates regardless of the noise level. However, we also find that our DAE based PSF model can further improve estimation accuracy when the noise level is high. These results show that our DAE PSF model can be used to increase performance of post-processing methods.\\

However, it should be noted that since there are some correlations between tilt and decenter, the estimation accuracy of these parameters is affected by these correlations. We have calculated correlation of errors between each predicted values as shown in table  \ref{tab:corref}. We find that decentX and tiltY, decentY and tiltY, and tiltX and tiltY have very strong positive correlations. Our DAE PSF model can not suppress these correlations. It is a problem and we will try to further discuss this problem in our future paper about the secondary mirror alignment method.\\

\begin{table}
	\caption{This table shows the mean and variance of errors between the estimated and the original values of tilt and decenter for original images and PSFs obtained by DAE PSF model with Possion noise of $\lambda=0.002$.}
	\setlength{\tabcolsep}{2mm}{
	\begin{tabular}{ccc}
		\hline
		$\sigma=0.002$ & mean value  & variance  \\
		\hline
		DAE PSF   &\makecell[l]{$2.28\times10^{-2}$,$2.65\times10^{-2}$,\\$4.07\times10^{-2}$,$1.74\times10^{-2}$}      
		                &\makecell[l]{$2.74\times10^{-4}$,$3.58\times10^{-4}$,\\$3.17\times10^{-4}$,$1.71\times10^{-4}$} \\     
		Original data      &\makecell[l]{$1.21\times10^{-1}$, $1.78\times10^{-1}$,\\$4.56\times10^{-2}$,$9.30\times10^{-2}$} 
		                &\makecell[l]{$4.20\times10^{-3}$,$6.05\times10^{-3}$,\\$1.01\times10^{-3}$,$2.43\times10^{-3}$}  \\
		\hline
	\end{tabular}}
	\label{tab:diff_0.002}
\end{table}

\begin{table}
	\caption{This table shows the mean and variance of errors between the estimated and the original values of tilt and decenter for original images and PSFs obtained by DAE PSF model with Possion noise of  $\lambda=0.005$.}
	\setlength{\tabcolsep}{2mm}{
	\begin{tabular}{ccc}
		\hline
		$\sigma=0.005$ & mean value  & variance \\
		\hline
		DAE PSF   &\makecell[l]{ $2.92\times10^{-2}$, $2.62\times10^{-2}$,\\$4.38\times10^{-2}$, $1.98\times10^{-2}$}        
		                &\makecell[l]{$2.81\times10^{-4}$, $3.78\times10^{-4}$,\\$3.87\times10^{-4}$, $2.07\times10^{-4}$} \\  
		Original data      &\makecell[l]{$6.76\times10^{-1}$,$8.02\times10^{-1}$,\\$1.39\times10^{-1}$,$7.32\times10^{-1}$ }          
		                &\makecell[l]{$2.36\times10^{-2}$, $3.58\times10^{-2}$,\\$1.62\times10^{-2}$, $1.49\times10^{-2}$}  \\ 
		\hline
	\end{tabular}}
	\label{tab:diff_0.005}
\end{table}

\begin{table}
	\caption{The table shows the correlation coefficients between estimation errors of different parameters.}

	\begin{tabular}{cccc}
	\hline
	 correlation coefficients  &decenterX-tiltY&decenterY-tiltY&tiltX-tiltY\\
	\hline

	Original Data 0.002   &0.3787&0.0937&0.1255\\
      Original Data 0.005   &0.2386&0.0128&0.1351\\
	DAE PSF 0.002         &0.1461&0.1798&0.3224\\
	DAE PSF 0.005         &0.1182&0.2098&0.3313\\
	\hline
	\end{tabular}
    \label{tab:corref}
\end{table}

\section{Conclusions and Future Work}
In this paper, we propose a DAE based PSF modelling method. Our method assumes the PSF can be represented by the PSF templates obtained by calibration data. According to real observation conditions, we train the DAE with PSF templates and simulated observation data. After training, the DAE can be used to map any star image to its original PSF. Our method can obtain the original PSF regardless of the noise level and random aberration interference. We find that our DAE based PSF model can increase the accuracy of telescope secondary mirror alignment. Our work shows that the state of a telescope, which represents by the PSF can be well described by a trained neural network. It provides a new approach in understanding the PSF of telescopes. In the future, we will design post-processing methods with the DAE based PSF model to further increase the observation data quality in WFSATs. Besides, obtaining the map between the shape of point spread function and their position in the field of view is also important. We will carry out our further research in this area in the future.

\section*{Acknowledgements}
The authors would like to thank the anonymous referee for comments and suggestions that greatly improved the quality of this manuscript. Peng Jia would like to thank Dr. Alastair Basden from Durham University, Dr. Rongyu Sun from Purple Mountain Observatory who provide very helpful suggestions for this paper. This work is supported by National Natural Science Foundation of China (NSFC)(11503018), the Joint Research Fund in Astronomy (U1631133) under cooperative agreement between the NSFC and Chinese Academy of Sciences (CAS),Shanxi Province Science Foundation for Youths (201901D211081), Research and Development Program of Shanxi (201903D121161), Research Project Supported by Shanxi Scholarship Council of China, the Scientific and Technological Innovation Programs of Higher Education Institutions in Shanxi (2019L0225).\\




\bibliographystyle{mnras}
\bibliography{DAE} 

\begin{thebibliography}{}
\makeatletter
\relax
\def\mn@urlcharsother{\let\do\@makeother \do\$\do\&\do\#\do\^\do\_\do\%\do\~}
\def\mn@doi{\begingroup\mn@urlcharsother \@ifnextchar [ {\mn@doi@}
  {\mn@doi@[]}}
\def\mn@doi@[#1]#2{\def\@tempa{#1}\ifx\@tempa\@empty \href
  {http://dx.doi.org/#2} {doi:#2}\else \href {http://dx.doi.org/#2} {#1}\fi
  \endgroup}
\def\mn@eprint#1#2{\mn@eprint@#1:#2::\@nil}
\def\mn@eprint@arXiv#1{\href {http://arxiv.org/abs/#1} {{\tt arXiv:#1}}}
\def\mn@eprint@dblp#1{\href {http://dblp.uni-trier.de/rec/bibtex/#1.xml}
  {dblp:#1}}
\def\mn@eprint@#1:#2:#3:#4\@nil{\def\@tempa {#1}\def\@tempb {#2}\def\@tempc
  {#3}\ifx \@tempc \@empty \let \@tempc \@tempb \let \@tempb \@tempa \fi \ifx
  \@tempb \@empty \def\@tempb {arXiv}\fi \@ifundefined
  {mn@eprint@\@tempb}{\@tempb:\@tempc}{\expandafter \expandafter \csname
  mn@eprint@\@tempb\endcsname \expandafter{\@tempc}}}

\bibitem[\protect\citeauthoryear{{Bailey}}{{Bailey}}{2012}]{Bailey2012}
{Bailey} S.,  2012, \mn@doi [\pasp] {10.1086/668105}, \href
  {https://ui.adsabs.harvard.edu/abs/2012PASP..124.1015B} {124, 1015}

\bibitem[\protect\citeauthoryear{{Basden}, {Bharmal}, {Jenkins}, {Morris},
  {Osborn}, {Peng}  \& {Staykov}}{{Basden} et~al.}{2018}]{Basden2018}
{Basden} A.~G.,  {Bharmal} N.~A.,  {Jenkins} D.,  {Morris} T.~J.,  {Osborn} J.,
   {Peng} J.,   {Staykov} L.,  2018, \mn@doi [SoftwareX]
  {10.1016/j.softx.2018.02.005}, \href
  {https://ui.adsabs.harvard.edu/abs/2018SoftX...7...63B} {7, 63}

\bibitem[\protect\citeauthoryear{Bourlard \& Kamp}{Bourlard \&
  Kamp}{1988}]{bourlard1988auto}
Bourlard H.,  Kamp Y.,  1988, Biological cybernetics, 59, 291

\bibitem[\protect\citeauthoryear{Burd et~al.,}{Burd et~al.}{2005}]{burd2005pi}
Burd A.,  et~al., 2005, in Photonics Applications in Industry and Research IV.
  p. 59481H

\bibitem[\protect\citeauthoryear{Cavallari, Ribeiro  \& Ponti}{Cavallari
  et~al.}{2018}]{cavallari2018unsupervised}
Cavallari G.,  Ribeiro L.,   Ponti M.,  2018, pp 440--446

\bibitem[\protect\citeauthoryear{Cha, Kim  \& Lee}{Cha
  et~al.}{2019}]{cha2019transformation}
Cha J.,  Kim K.~S.,   Lee S.,  2019, arXiv preprint arXiv:1901.08479

\bibitem[\protect\citeauthoryear{Grimm \& Heng}{Grimm \&
  Heng}{2015}]{grimm2015helios}
Grimm S.~L.,  Heng K.,  2015, The Astrophysical Journal, 808, 182

\bibitem[\protect\citeauthoryear{Hotelling}{Hotelling}{1933}]{hotelling1933analysis}
Hotelling H.,  1933, Journal of Educational Psychology, 24, 417

\bibitem[\protect\citeauthoryear{Ichimura}{Ichimura}{2018}]{ichimura2018spatial}
Ichimura N.,  2018, arXiv preprint arXiv:1806.02336

\bibitem[\protect\citeauthoryear{Ioffe \& Szegedy}{Ioffe \&
  Szegedy}{2015}]{ioffe2015batch}
Ioffe S.,  Szegedy C.,  2015, arXiv preprint arXiv:1502.03167

\bibitem[\protect\citeauthoryear{Jee \& Tyson}{Jee \&
  Tyson}{2011}]{jee2011toward}
Jee M.~J.,  Tyson J.~A.,  2011, Publications of the Astronomical Society of the
  Pacific, 123, 596

\bibitem[\protect\citeauthoryear{Jee, Blakeslee, Sirianni, Martel, White  \&
  Ford}{Jee et~al.}{2007}]{jee2007principal}
Jee M.,  Blakeslee J.,  Sirianni M.,  Martel A.,  White R.,   Ford H.,  2007,
  Publications of the Astronomical Society of the Pacific, 119, 1403

\bibitem[\protect\citeauthoryear{Jia, Cai, Wang  \& Basden}{Jia
  et~al.}{2015a}]{jia2015simulation}
Jia P.,  Cai D.,  Wang D.,   Basden A.,  2015a, Monthly Notices of the Royal
  Astronomical Society, 447, 3467

\bibitem[\protect\citeauthoryear{Jia, Cai, Wang  \& Basden}{Jia
  et~al.}{2015b}]{jia2015real}
Jia P.,  Cai D.,  Wang D.,   Basden A.,  2015b, Monthly Notices of the Royal
  Astronomical Society, 450, 38

\bibitem[\protect\citeauthoryear{{Jia}, {Sun}, {Wang}, {Cai}  \& {Liu}}{{Jia}
  et~al.}{2017}]{Jia2017}
{Jia} P.,  {Sun} R.,  {Wang} W.,  {Cai} D.,   {Liu} H.,  2017, \mn@doi [\mnras]
  {10.1093/mnras/stx1336}, \href
  {https://ui.adsabs.harvard.edu/abs/2017MNRAS.470.1950J} {470, 1950}

\bibitem[\protect\citeauthoryear{Kingma \& Ba}{Kingma \&
  Ba}{2014}]{kingma2014adam}
Kingma D.~P.,  Ba J.,  2014, arXiv preprint arXiv:1412.6980

\bibitem[\protect\citeauthoryear{Kohonen}{Kohonen}{1982}]{kohonen1982self}
Kohonen T.,  1982, Biological cybernetics, 43, 59

\bibitem[\protect\citeauthoryear{Kossaifi, Panagakis, Anandkumar  \&
  Pantic}{Kossaifi et~al.}{2019}]{kossaifi2019tensorly}
Kossaifi J.,  Panagakis Y.,  Anandkumar A.,   Pantic M.,  2019, The Journal of
  Machine Learning Research, 20, 925

\bibitem[\protect\citeauthoryear{Laurent \& von Brecht}{Laurent \& von
  Brecht}{2017}]{laurent2017multilinear}
Laurent T.,  von Brecht J.,  2017, arXiv preprint arXiv:1712.10132

\bibitem[\protect\citeauthoryear{{Li}, {Yuan}  \& {Cui}}{{Li}
  et~al.}{2015}]{Li2015}
{Li} Z.,  {Yuan} X.,   {Cui} X.,  2015, \mn@doi [\mnras]
  {10.1093/mnras/stv268}, \href
  {https://ui.adsabs.harvard.edu/abs/2015MNRAS.449..425L} {449, 425}

\bibitem[\protect\citeauthoryear{{Li}, {Li}, {Cheng}, {Peterson}  \&
  {Cui}}{{Li} et~al.}{2016}]{Li2016}
{Li} B.-S.,  {Li} G.-L.,  {Cheng} J.,  {Peterson} J.,   {Cui} W.,  2016,
  \mn@doi [Research in Astronomy and Astrophysics]
  {10.1088/1674-4527/16/9/139}, \href
  {https://ui.adsabs.harvard.edu/abs/2016RAA....16..139L} {16, 139}

\bibitem[\protect\citeauthoryear{{Ma}, {Zhao}  \& {Yao}}{{Ma}
  et~al.}{2007}]{Ma2007}
{Ma} Y.,  {Zhao} H.,   {Yao} D.,  2007, in {Valsecchi} G.~B.,
  {Vokrouhlick{\'y}} D.,   {Milani} A.,  eds,  IAU Symposium Vol. 236, Near
  Earth Objects, our Celestial Neighbors: Opportunity and Risk. pp 381--384,
  \mn@doi{10.1017/S1743921307003468}

\bibitem[\protect\citeauthoryear{Makidon, Casertano, Cox  \& van~der
  Marel}{Makidon et~al.}{2007}]{makidon2007jwst}
Makidon R.,  Casertano S.,  Cox C.,   van~der Marel R.,  2007, NASA Technic Al
  Report

\bibitem[\protect\citeauthoryear{Moffat}{Moffat}{1969}]{moffat1969theoretical}
Moffat A.,  1969, Astronomy and Astrophysics, 3, 455

\bibitem[\protect\citeauthoryear{Perrin, Sivaramakrishnan, Lajoie, Elliott,
  Pueyo, Ravindranath  \& Albert}{Perrin et~al.}{2014}]{perrin2014updated}
Perrin M.~D.,  Sivaramakrishnan A.,  Lajoie C.,  Elliott E.,  Pueyo L.,
  Ravindranath S.,   Albert L.,  2014, Proceedings of SPIE, 9143

\bibitem[\protect\citeauthoryear{{Perrin}, {Long}, {Douglas},
  {Sivaramakrishnan}  \& {Slocum}}{{Perrin} et~al.}{2016}]{Perrin2016}
{Perrin} M.,  {Long} J.,  {Douglas} E.,  {Sivaramakrishnan} A.,   {Slocum} C.,
  2016, {POPPY: Physical Optics Propagation in PYthon} (\mn@eprint {ascl}
  {1602.018})

\bibitem[\protect\citeauthoryear{Ping \& Zhang}{Ping \&
  Zhang}{2017}]{ping2017the}
Ping Y.,  Zhang C.,  2017, Advances in Space Research, 60, 907

\bibitem[\protect\citeauthoryear{Piotrowski et~al.,}{Piotrowski
  et~al.}{2013}]{piotrowski2013psf}
Piotrowski L.~W.,  et~al., 2013, Astronomy and Astrophysics, 551, A119

\bibitem[\protect\citeauthoryear{Racine}{Racine}{1996}]{racine1996telescope}
Racine R.,  1996, Publications of the Astronomical Society of the Pacific, 108,
  699

\bibitem[\protect\citeauthoryear{Ratzloff, Law, Fors, Corbett, Howard, Ser  \&
  Haislip}{Ratzloff et~al.}{2019}]{ratzloff2019building}
Ratzloff J.~K.,  Law N.~M.,  Fors O.,  Corbett H.~T.,  Howard W.~S.,  Ser
  D.~D.,   Haislip J.~B.,  2019, Publications of the Astronomical Society of
  the Pacific, 131, 075001

\bibitem[\protect\citeauthoryear{Rhodes, Massey, Albert, Taylor, Koekemoer  \&
  Leauthaud}{Rhodes et~al.}{2005}]{rhodes2005modeling}
Rhodes J.,  Massey R.,  Albert J.,  Taylor J.~E.,  Koekemoer A.~M.,   Leauthaud
  A.,  2005, arXiv: Astrophysics

\bibitem[\protect\citeauthoryear{Rhodes et~al.,}{Rhodes
  et~al.}{2007}]{rhodes2007stability}
Rhodes J.~D.,  et~al., 2007, The Astrophysical Journal Supplement Series, 172,
  203

\bibitem[\protect\citeauthoryear{Sun \& Jia}{Sun \& Jia}{2017}]{sun2017image}
Sun R.,  Jia P.,  2017, Publications of the Astronomical Society of the
  Pacific, 129, 044502

\bibitem[\protect\citeauthoryear{Sun \& Yu}{Sun \& Yu}{2019}]{sun2019precise}
Sun R.,  Yu S.,  2019, Astrophysics and Space Science, 364, 39

\bibitem[\protect\citeauthoryear{Vidal, Ma  \& Sastry}{Vidal
  et~al.}{2005}]{vidal2005generalized}
Vidal R.,  Ma Y.,   Sastry S.~S.,  2005, IEEE Transactions on Pattern Analysis
  and Machine Intelligence, 27, 1945

\bibitem[\protect\citeauthoryear{Vincent, Larochelle, Bengio  \&
  Manzagol}{Vincent et~al.}{2008}]{vincent2008extracting}
Vincent P.,  Larochelle H.,  Bengio Y.,   Manzagol P.-A.,  2008, pp 1096--1103

\bibitem[\protect\citeauthoryear{Vincent, Larochelle, Lajoie, Bengio  \&
  Manzagol}{Vincent et~al.}{2010}]{vincent2010stacked}
Vincent P.,  Larochelle H.,  Lajoie I.,  Bengio Y.,   Manzagol P.-A.,  2010,
  Journal of Machine Learning Research, 11, 3371

\bibitem[\protect\citeauthoryear{Wang, Jia, Cai  \& Liu}{Wang
  et~al.}{2018}]{wang2018automated}
Wang W.,  Jia P.,  Cai D.,   Liu H.,  2018, Monthly Notices of the Royal
  Astronomical Society, 478, 5671

\bibitem[\protect\citeauthoryear{{van der Walt} et~al.,}{{van der Walt}
  et~al.}{2014}]{van2014}
{van der Walt} S.,  et~al., 2014, arXiv e-prints, \href
  {https://ui.adsabs.harvard.edu/abs/2014arXiv1407.6245V} {p. arXiv:1407.6245}

\makeatother
\end{thebibliography}








\bsp	
\label{lastpage}
\end{document}